# Novel Binary Addition Tree Algorithm (BAT) for Calculating the Direct Lower-Bound of the Highly Reliable Binary-State Network Reliability

Wei-Chang Yeh
Department of Industrial Engineering and Engineering Management
National Tsing Hua University
P.O. Box 24-60, Hsinchu, Taiwan 300, R.O.C.
yeh@ieee.org

*Abstract* — Real-world applications such as the internet of things, wireless sensor networks, smart grids, transportation networks, communication networks, social networks, and computer grid systems are typically modeled as network structures. Network reliability represents the success probability of a network and it is an effective and popular metric for evaluating the performance of all types of networks. Binary-state networks composed of binary-state (e.g., working or failed) components (arcs and/or nodes) are some of the most popular network structures. The scale of networks has grown dramatically in recent years. For example, social networks have more than a billion users. Additionally, the reliability of components has increased as a result of both mature and emergent technology. For highly reliable networks, it is more practical to calculate approximated reliability, rather than exact reliability, which is an NP-hard problem. Therefore, we propose a novel direct reliability lower bound based on the binary addition tree algorithm to calculate approximate reliability. The efficiency and effectiveness of the proposed reliability bound are analyzed based on time complexity and validated through numerical experiments.

## 1. INTRODUCTION

Binary-state networks are essential structures in various real-world network applications, including communication [1], distribution [2, 3, 19], transportation [4, 17], transformation [5, 21], transmission, recovery, and backup networks for power [6], signals [7, 22], liquids, gases [8, 16], data [9, 10], multimedia [11, 12], topology design [13, 14, 20], and resilience [15, 18]. Therefore, binary-state networks have attracted significant attention for research and applications in the planning, design, execution, management, and control of all the aforementioned systems.

A network functions properly if source nodes and sink nodes can communicate to send signals, power, data, etc. The reliability of a binary-state network is defined as the probability that the network will function properly and that at least one directed path can be found in the network. Reliability is a very useful performance and functionality metric for networks. Reliability has been widely used in many studies in recent decades [6, 16–21, 23–26].

For example, Aven proposed a standard for measuring the reliability of binary-state transportation networks [16]. Bhavathrathan and Patil proposed a model to determine the worst state of link capacity in virtual scenarios to evaluate the reliability of binary-state transportation networks [17]. Kakadia and Ramirez-Marquez considered resiliency based on the reliability of a binary-state telecom network [18]. Laitrakun and Coyle utilized a splitting algorithm to calculate the reliability of a binary-state wireless sensor network under time constraints [19]. Lin et al. designed a network topology to optimize the reliability of binary-state wireless sensor networks [20]. Ramirez-Marquez introduced an unscented transformation method for an uncertainty binary-state network [21]. Yeh adopted a squeezed artificial neural network and response surface methodology to construct binary-state network symbolic reliability functions [23, 24]. Yeh et al. developed swarm algorithms to optimize binary-state network reliability in the internet of things [25] and in a grid network [26]. Zhang et al. proposed a novel energy technology to evaluate binary-state grid network reliability [6].

Calculating the exact reliability of a binary-state network is an NP-hard problem [27, 28, 29]. Various tools have been proposed to estimate binary-state network reliability [30, 31], including exact reliability and approximate reliability methods. Exact solution methods focus on calculating exact reliability and can be divided into two main categories: indirect exact reliability methods and direct exact reliability methods.

Indirect exact solution methods must determine all minimal cuts (MCs) [31–35] or minimal paths (MPs) [36–39] and then implement the sum-of-disjoint method or inclusion-exclusion method to calculate the reliability in terms of the identified MCs or MPs. These two steps are both NP-hard. Direct methods such as the state-space algorithm, binary-decision diagram [40–41], and the binary addition tree algorithm (BAT) [42–44] can calculate reliability directly. In general, a direct algorithm is more efficient than an indirect algorithm. However, the binary-decision diagram requires extensive coding expertise using massive data structures [45]. Therefore, prior to the invention of the BAT, indirect algorithms based on MCs or MPs were more popular than direct algorithms [31, 33, 34, 36, 37, 38, 39]. However, the BAT is simple to code, flexible, and more efficient than indirect algorithms. Therefore, the BAT can be extended to different types of network reliability problems [42–44] and implemented to solve various practical problems.

The main advantage of the approximate reliability method is that it can overcome the obstacle of NP-hardness by finding a good solution, rather than an exact reliability value. There are also many approximate reliability methods that can be categorized as direct approximate reliability methods and indirect approximate reliability methods. Direct approximate reliability methods are mainly based on Monte Carlo simulation (MCS) [46, 47] and include all MCS-based algorithms [48, 49, 50].

Similar to direct exact reliability algorithms, direct approximate reliability methods can calculate approximate reliability without utilizing MPs or MCs. The fundamental concept of MCS-based algorithms is to generate a uniform random number for each variable to derive multiple results for a certain number of replications and then average the results to obtain an approximate reliability value. Existing MCS-based algorithms face a common problem in that they cannot provide a fixed approximate reliability value based on the inclusion of random variables. Additionally, MCS-based algorithms cannot determine whether the obtained approximate reliability is larger or smaller than the exact reliability.

Indirect approximate reliability algorithms focus on finding reliability bounds and they are very similar to indirect exact reliability algorithms in that they first find some MPs or MCs and then implement the sum-of-disjoint method or inclusion-exclusion method in terms of these MPs or MCs. The only difference between indirect exact reliability algorithms and indirect approximate reliability algorithms is that the former require all MPs or MCs and the latter can have a smaller number of MPs or MCs. The greater the number of MPs or MCs used in indirect approximate reliability algorithms, the closer they come to the exact reliability. However, as mentioned previously, using all MPs or MCs and implementing the sum-of-disjoint method or inclusion-exclusion method is NP-hard.

Therefore, it is either too difficult to develop an easy method to derive closer bounds or too tedious and inefficient to use existing methods to derive close bounds.

Based on the above discussion of existing reliability algorithms, we can draw the following conclusions.

1. Each exact reliability method, regardless of whether it is indirect or direct, is limited by the size of the network as a result of NP-hardness.
2. Direct approximate reliability methods are only based on MCS and are weak at predicting approximated reliability.
3. Indirect approximate reliability methods, which mainly find reliability bounds based on MPs/MCs, are less efficient at obtaining an accurate approximated reliability.

Real-world network applications such as grid computing, social networks, and the internet of things are becoming broader and more flexible [51]. The scale of practical networks is also growing over time. Therefore, there is an increasing need to develop more efficient algorithms to calculate acceptable approximate reliability [31, 51]. Additionally, in the current market environment, the lifecycle of electronic products is becoming shorter. Research and development time have been significantly reduced and the use of effective, low-cost, and highly reliable components has become the key to product design.

MP/MC-based algorithms are the main indirect exact reliability and indirect approximate reliability algorithms. However, the authors of [42–44] revealed that MP/MC-based algorithms are less efficient than BAT for evaluating the reliability of a binary-state network. Therefore, the goal of our study was to develop a simple and direct approximate reliability method called AppBAT based on the BAT to obtain a predictable, acceptable, and bounded reliability for highly reliable binary-state networks without implementing MCS or finding any MPs/MCs.

The remainder of this paper is organized as follows. Section 2 presents the acronyms, notations, nomenclatures, and assumptions adopted in this study. Section 3 reviews the fundamental components of the proposed AppBAT, including the BAT and path-based layered search algorithm (PLSA). A novel implicit BAT is proposed to find each vector that has a constant number of coordinates with a value of one in Section 4. The proposed AppBAT based on the implicit BAT is presented in Section 5 with pseudocode, a step-by-step example, and time complexity analysis. In addition, a computational experiment was performed to analyze the relationships among approximate reliability, arc reliability, and $z$, where $z$ is the minimal number of coordinates with a value of one for the connected vectors used to calculate approximate reliability. Finally, Section 6 concludes this paper.

## 2. ACRONYMS, NOTATIONS, NOMENCLATURES, AND ASSUMPTIONS

Important acronyms, notations, assumptions, and nomenclatures related to the proposed algorithm are presented in this section.

### 2.2 Notations

$a_i$: arc $i$ in $E$ for $i = 0, 1, \ldots, m-1$

$\Pr(\bullet)$: occurrence probability of event $\bullet$

$\Pr_p(\bullet)$: $\Pr(\bullet) = p$

**D**: arc state distribution including the states and related success probabilities of each arc

$G(V, E)$: an undirected graph with $V$ and $E$ (e.g., in Fig. 1, a graph with $V = \{1, 2, 3, 4\}$ and $E = \{a_0, a_1, a_2, a_3, a_4\}$, where the source node is 1 and the sink node is 4

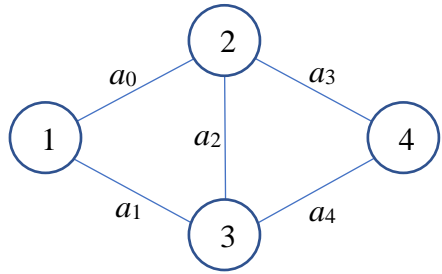

**Figure 1.** Example network

$G(V, E, \mathbf{D})$: a binary-state network with an undirected graph $G(V, E)$ and binary-state distributions **D** (e.g., Fig. 1 is a binary-state network and **D** is given in Table 1)

**Table 1.** Binary-state distribution **D** in Fig. 1

| $i$ | $\Pr(a_i)$ |
|---|---|
| 0 | 0.8 |
| 1 | 0.9 |
| 2 | 0.7 |
| 3 | 0.8 |

$\Xi_z$:    $z$-tuple (one-based) indicating vector, where $\Xi_z(i) = j$ if and only if $X(a_j) = 1$, where $X$ is the corresponding state vector of $\Xi_z$

$\Xi_z^*$:    $z$-tuple (zero-based) indicating vector, where $\Xi_z^*(i) = j$ if and only if $X(a_j) = 0$, where $X$ is the corresponding state vector of $\Xi_z^*$

$X$:    ($m$-tuple) state vector obtained from the forward BAT

$\underline{X}$:    ($m$-tuple) state vector obtained from the backward BAT

$X(a_i)$:    value of the coordinate $i$ for $i = 0, 1, \ldots, (m\text{-}1)$ (i.e., $a_i$, of $X$, $X(a_2) = X(a_4) = X(a_5) =1$ if $X = (0, 1, 0, 1, 1)$)

Zero($X$):    number of coordinates with values of zero (e.g., Zero($X$) = 2 if $X = (0, 1, 0, 1, 1)$)

One($X$):    One($X$) = $\sum_{i=0}^{m-1} X(a_i)$ (e.g., One($X$) = 3 if $X = (0, 1, 0, 1, 1)$)

$W_f(X)$: $W_f(X) = \sum_{i=0}^{m-1} 2^i X(a_i)$

$W_b(X)$: $W_b(X) = \sum_{i=0}^{m-1} 2^{m-i-1} X(a_i)$

T(•):    runtime to have event •

Pr(•):    probability to have event •

$\Pr_p(\bullet)$:    probability to have event • when $\Pr(a) = p$ for all $a \in E$

Pr($X$):    Pr($X$) = Pr($X(a_1)$)·Pr($X(a_2)$)·…·Pr($X(a_m)$) if $X$ is connected

$\Pr_p(X)$:    $\Pr_p(X) = \Pr_p(X(a_1))\cdot\Pr_p(X(a_2))\cdot\ldots\cdot\Pr_p(X(a_m))$ if $X$ is connected

$R(G)$:    reliability of $G(V, E, \mathbf{D})$

$R_z$:    approximate reliability of $G(V, E, \mathbf{D})$ in terms of the connected vectors $X$ with $z \leq$ One($X$)

$G(X)$:    subgraph represented by the state vector $X$ such that $G(X) = G(V, \{a \in E \mid$ for all $a$ with $X(a) = 1\})$ (e.g., the graph of $G(X)$ is depicted in Fig. 2, where $X = (1, 1, 1, 0, 0)$)

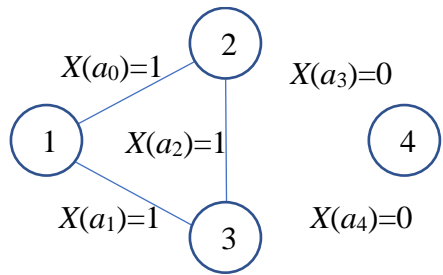

**Figure 2.** $G(X)$ and $X = (1, 1, 1, 0, 0)$ from Fig. 1.

$A \ll B$: vector $A$ is obtained earlier than vector $B$ in the procedures of BAT-based algorithms

$\overline{\Omega}_z$: $\overline{\Omega}_z = \{X \mid X \text{ is a state vector and One}(X) = z\}$

$\Omega_z$: $\Omega_z = \{X \mid X \text{ is connected and One}(X) = z\}$

### 2.3 Nomenclatures

Reliability: The success probability that a network is functioning.

Connected vector: A state vector $X$ is connected if nodes 1 and $n$ are connected by at least one directed path.

MP: A simple path from source node 1 to sink node $n$ is an MP. For example, $\{a_0, a_2, a_4\}$ is an MP from node 1 to node $n$, as shown in Fig. 1. Note that the removal of any arc from a simple path will cause the residual simple path to no longer be a path.

MC: A cut between source node 1 and sink node $n$ is an MC if any proper subset of an MC is not an MC. For example, $\{a_0, a_1\}$ is an MC in Fig. 1.

### 2.4 Assumptions

1. All nodes are completely reliable and connected in networks.
2. Each arc has two states in a network and its probability of success is statistically independent according to a given distribution.
3. A network has no parallel arcs or loops.

### 3. REVIEW OF THE BAT AND PLSA

The proposed implicit BAT and AppBAT change the sequence of finding vectors in the BAT to improve the efficiency of the BAT for calculating the approximate reliability of binary-state networks. The PLSA [52] is a common method used to verify the connectivity of obtained vectors in the proposed AppBAT for calculating approximate reliability. Therefore, both the BAT [42–44] and PLSA [52] are briefly reviewed in this section.

## 3.1 BAT

The BAT proposed by Yeh can generate all state vectors using only the binary addition operator. In a state vector of a binary-state network, the coordinates are either zero or one, indicating that the corresponding arcs fail or work, respectively []. The BAT simply lists all $m$-tuple state vectors using a procedure analogous to adding one to a binary code.

The backward BAT adds one to the last coordinate in the binary-state vector and gradually moves to the first coordinate [42–44]. In contrast, the forward BAT adds one to the first coordinate in the binary-state vector and gradually moves to the last coordinate [42–44]. The overall procedure of the binary-addition tree for the forward BAT is described by the pseudocode below.

**Algorithm: Forward BAT**

**Input:** $G(V, E)$.

**Output:** All state vectors without duplications.

**STEP B0.** Let $i = \text{SUM} = 0$ and $X = \mathbf{0}$.

**STEP B1.** If $X(a_i) = 0$, let $X(a_i) = 1$, $\text{SUM} = \text{SUM} + 1$, and go to STEP B3.

**STEP B2.** If $i > 1$, let $X(a_i) = 0$, $\text{SUM} = \text{SUM} - 1$, $i = i + 1$, and go to STEP B1.

**STEP B3.** If $\text{SUM} = m$, halt. Otherwise, let $i = 1$ and go to STEP B1.

In STEP B0, we initialize the values of the current coordinate $i$ to the first coordinate (i.e., coordinate zero), the first state vector and current vector to vector zero, and $\text{SUM} = 0$, which is the number of coordinates in $X$. The loop from STEP B1 to STEP B3 finds each state vector iteratively by adding one to the current coordinate of the current state vector (e.g., $X(a_i)$) after considering $X$ as a binary number such that its digit $i$ is equal to $X(a_i)$ for $i = 0, 1, \ldots, (m - 1)$. This loop terminates when the number of ones in $X$ is $m$, meaning that $\text{SUM} = m$ and $X$ is the vector.

For the backward BAT, "$i = 0$" and "$i = i + 1$" in STEP B0 and STEP B2 in the forward BAT are changed to "$i = m - 1$" and "$i = i - 1$," respectively.

The time complexities of both the forward and backward BAT are ($2^m + 1$), as proven in [42–44]. There are various implementations of the BAT [42–44] because it is easy to code, flexible, and more efficient than other search methods such as depth-first search and breadth-first search. Let $X$ and $Y$ be two vectors obtained from the BAT. Then, there is one commonality shared by all BAT implementations.

For all forward BAT, we have

$$W_f(X) = \sum_{i=0}^{m-1} 2^i X(a_i) < W_f(Y) = \sum_{i=0}^{m-1} 2^i Y(a_i) \text{ if and only if } X \ll Y, \quad (1)$$

and for all backward BAT, we have

$$W_b(X) = \sum_{i=0}^{m-1} 2^{m-i+1} X(a_i) < W_b(Y) = \sum_{i=0}^{m-1} 2^{m-i+1} Y(a_i) \text{ if and only if } X \ll Y. \quad (2)$$

For example, in Fig. 1, each state vector is a 5-tuple and $X = (0, 0, 0, 0, 0)$ is the first vector. From the binary addition in the backward BAT, we have the second vector $(0, 0, 0, 0, 1)$, third vector $(0, 0, 0, 1, 0)$, …, and last vector $(1, 1, 1, 1, 1)$. Similarly, we have all of the state vectors. Table 2 lists the values of each state vector (e.g., $X$) obtained from the forward BAT with $W_f(X)$, Zero($X$), One($X$), connectivity of $X$, $Pr_{0.9}(X)$ if $X$ is connected, and the same sequence order of the state vector $\underline{X}$ obtained from the backward BAT. For example, $X_{14} = (1, 0, 1, 1, 0)$ is obtained in the 14$^{th}$ iteration from the forward BAT, $W_f(X) = 14$, Zero($X_{14}$) = 2, One($X_{14}$) = 3, $X_{14}$ is connected with $Pr_{0.9}(X_{14}) = 0.00729$, and the 14$^{th}$ vector obtained from the backward BAT is $\underline{X}_{14} = (0, 1, 1, 0, 1)$.

It should be noted that $X(a_i) = \underline{X}(a_{m-i+1})$ for $i = 0, 1, …, (m-1)$, $W_f(X) = W_b(\underline{X})$, Zero($X$) = Zero($\underline{X}$), One($X$) = One($\underline{X}$), and Zero($X$) + One($X$) = $m$.

**Table 2.** All vectors obtained from the BAT.

| i | $X_i$ | $W_f(X_i)$ | Zero($X_i$) | One($X_i$) | Connected? | $Pr_{0.9}(X_i)$ | $\underline{X}_i$ |
|---|---|---|---|---|---|---|---|
| 1 | (0, 0, 0, 0, 0) | 0 | 5 | 0 | N | 0 | (0, 0, 0, 0, 0) |
| 2 | (1, 0, 0, 0, 0) | 1 | 4 | 1 | N | 0 | (0, 0, 0, 0, 1) |
| 3 | (0, 1, 0, 0, 0) | 2 | 4 | 1 | N | 0 | (0, 0, 0, 1, 0) |
| 4 | (1, 1, 0, 0, 0) | 3 | 3 | 2 | N | 0 | (0, 0, 0, 1, 1) |
| 5 | (0, 0, 1, 0, 0) | 4 | 4 | 1 | N | 0 | (0, 0, 1, 0, 0) |
| 6 | (1, 0, 1, 0, 0) | 5 | 3 | 2 | N | 0 | (0, 0, 1, 0, 1) |
| 7 | (0, 1, 1, 0, 0) | 6 | 3 | 2 | N | 0 | (0, 0, 1, 1, 0) |
| 8 | (1, 1, 1, 0, 0) | 7 | 2 | 3 | N | 0 | (0, 0, 1, 1, 1) |
| 9 | (0, 0, 0, 1, 0) | 8 | 4 | 1 | N | 0 | (0, 1, 0, 0, 0) |
| 10 | (1, 0, 0, 1, 0) | 9 | 3 | 2 | Y | 0.000810 | (0, 1, 0, 0, 1) |
| 11 | (0, 1, 0, 1, 0) | 10 | 3 | 2 | N | 0 | (0, 1, 0, 1, 0) |
| 12 | (1, 1, 0, 1, 0) | 11 | 2 | 3 | Y | 0.007290 | (0, 1, 0, 1, 1) |
| 13 | (0, 0, 1, 1, 0) | 12 | 3 | 2 | N | 0 | (0, 1, 1, 0, 0) |
| 14 | (1, 0, 1, 1, 0) | 13 | 2 | 3 | Y | 0.007290 | (0, 1, 1, 0, 1) |
| 15 | (0, 1, 1, 1, 0) | 14 | 2 | 3 | N | 0 | (0, 1, 1, 1, 0) |
| 16 | (1, 1, 1, 1, 0) | 15 | 1 | 4 | Y | 0.065610 | (0, 1, 1, 1, 1) |
| 17 | (0, 0, 0, 0, 1) | 16 | 4 | 1 | N | 0 | (1, 0, 0, 0, 0) |
| 18 | (1, 0, 0, 0, 1) | 17 | 3 | 2 | N | 0 | (1, 0, 0, 0, 1) |
| 19 | (0, 1, 0, 0, 1) | 18 | 3 | 2 | Y | 0.000810 | (1, 0, 0, 1, 0) |
| 20 | (1, 1, 0, 0, 1) | 19 | 2 | 3 | Y | 0.007290 | (1, 0, 0, 1, 1) |
| 21 | (0, 0, 1, 0, 1) | 20 | 3 | 2 | N | 0 | (1, 0, 1, 0, 0) |
| 22 | (1, 0, 1, 0, 1) | 21 | 2 | 3 | Y | 0.007290 | (1, 0, 1, 0, 1) |
| 23 | (0, 1, 1, 0, 1) | 22 | 2 | 3 | Y | 0.007290 | (1, 0, 1, 1, 0) |
| 24 | (1, 1, 1, 0, 1) | 23 | 1 | 4 | Y | 0.065610 | (1, 0, 1, 1, 1) |
| 25 | (0, 0, 0, 1, 1) | 24 | 3 | 2 | N | 0 | (1, 1, 0, 0, 0) |
| 26 | (1, 0, 0, 1, 1) | 25 | 2 | 3 | Y | 0.007290 | (1, 1, 0, 0, 1) |
| 27 | (0, 1, 0, 1, 1) | 26 | 2 | 3 | Y | 0.007290 | (1, 1, 0, 1, 0) |
| 28 | (1, 1, 0, 1, 1) | 27 | 1 | 4 | Y | 0.065610 | (1, 1, 0, 1, 1) |
| 29 | (0, 0, 1, 1, 1) | 28 | 2 | 3 | N | 0 | (1, 1, 1, 0, 0) |
| 30 | (1, 0, 1, 1, 1) | 29 | 1 | 4 | Y | 0.065610 | (1, 1, 1, 0, 1) |
| 31 | (0, 1, 1, 1, 1) | 30 | 1 | 4 | Y | 0.065610 | (1, 1, 1, 1, 0) |
| 32 | (1, 1, 1, 1, 1) | 31 | 0 | 5 | Y | 0.590490 | (1, 1, 1, 1, 1) |
| | SUM | | | | | 0.971190 | |

### 3.2 PLSA

The connectivity of each state vector obtained from the BAT must be validated before evaluating network reliability. To accomplish this goal effectively, the PLSA proposed in [52] is implemented to verify the connectivity of each sub-network represented by the state vector in the BAT. For example, the sub-networks represented by $X_8 = (1, 1, 1, 0, 0)$ are depicted in Fig. 2.

The PLSA was adapted from the layered-search algorithm (LSA), which was proposed in [52], to find all *d*-MPs in acyclic networks. Based on the efficiency and simplicity of the LSA, the LSA was revised into the PLSA to verify the connectivity of a state vector (e.g., *X*) by finding a direct path from source node 1 to the sink node in *G*(*X*) [52]. If such a path exists, then *X* is connected. Otherwise, *X* is disconnected.

Let $L_i$ be the layer *i*, $L_1 = \{1\}$, and $L_i = \{ v \in V \mid$ there is an arc from a node in $L_{i-1}$ to *v*, and no arc from $L_{i-k}$ to *v* for $k = 1, 2, …, i$ in $G(X)\}$. Because there are at most *n* nodes with at least one node in each layer, there are at most *n* layers. If $n \in L_j$ for some *j*, then *X* is connected. Otherwise, *X* is disconnected. Based on this simple concept proposed in the LSA [52], the PLSA can determine whether nodes 1 and *n* are connected in the subgraph related to the state vector to test the connectivity of the vector. The pseudocode for this procedure is presented below.

**Algorithm: PLSA**

**Input:** A state vector $X$.

**Output:** $X$ is connected or disconnected.

**STEP P0.** Let $L_1 = \{1\}$ and $i = 2$.

**STEP P1.** Let $L_i = \{ v \in V \mid$ there is an arc from a node in $L_{i-1}$ to $v$, and no arc from $L_{i-k}$ to $v$ for $k = 1, 2, \ldots, i$ in $G(X)\}$.

**STEP P2.** If $n \in L_i$, $X$ is connected and halt.

**STEP P3.** If $L_i = \emptyset$, $X$ is disconnected and halt. Otherwise, go to STEP P1.

There is at least one node in each layer and we have at most $n$ layers. Therefore, the time complexity of the PLSA is $O(n)$ for verifying whether a state vector is connected.

For example, based on Fig. 2, the PLSA procedure used to determine the connectivity of $X = (1, 1, 1, 0, 0)$ is presented in Table 3.

**Table 3.** Example PLSA procedure.

| $i$ | $L_i$ | Remark |
|---|---|---|
| 1 | $\{1\}$ | |
| 2 | $\{2, 3\}$ | |
| 3 | $\emptyset$ | $X$ is disconnected |

## 4 PROPOSED IMPLICIT BAT FOR GENERATING ALL X WITH ZERO(X) = z

A novel implicit BAT is proposed here for each vector $X$ with One$(X) = z$. The proposed AppBAT is based on the implicit BAT, which finds all vectors (e.g., $X$) with $z \leq$ One$(X)$ and calculates the approximate reliability in terms of the identified vectors.

### 4.1 Indicator Vector

In the traditional BAT, the only way to generate all vectors $X$ with One$(X) = z$ (i.e., Zero$(X) = (m - z)$) is to collect all vectors and then discard vectors (e.g., $Y$) with One$(Y) \neq z$. A novel BAT called the implicit BAT is proposed to solve the problem of generating all vectors $X$ with One$(X) = z$ by using another BAT.

Let the indicator vector $\Xi_z$ be a $z$-tube vector, where none of the values of its coordinates are identical. For example, $(0, 0)$ and $(1, 1)$ are not allowed in $\Xi_2$. Without loss of generality, let the $i$th coordinate in the state vector $X$ be the coordinate $(i - 1)$, for example, the first coordinate is called coordinate zero in $X$. The value of $\Xi_z(i)$ indicates whether the value of coordinate $i$ is one in the state vector $X$. For example, let $z = 2$. Then, $\Xi_z = (1, 2)$ indicates that the positions of the ones are at coordinates one and two in $X$ (i.e., (0, 1, 1, 0, 0)). Hence, there is a one-to-one relationship between $\Xi_z$ and $X$.

**4.2 Pseudocode for the Proposed Implicit BAT**

The pseudocode for the proposed implicit BAT is presented below.

**Algorithm: Implicit BAT**

**Input:**   $G(V, E)$.

**Output:**   All vectors $X$ with One($X$) = $z$.

**STEP I0.**   Let $i = 0$, $\Xi_z = (0, 1, 2, \dots, z-1)$, and $\Xi_z(z) = m$. Note that the original $\Xi_z$ is a $z$-tuple vector.

**STEP I1.**   Based on $\Xi_z$, find the corresponding $X$ such that $X(a_j) = 1$ if $\Xi_z(i) = j$ for one $i$ and $X(a_j) = 0$ if $\Xi_z(i) \neq j$ for all $i$.

**STEP I2.**   If $\Xi_z(i) < \Xi_z(I + 1) - 1$, then let $\Xi_z(i) = \Xi_z(i) + 1$, $i = 0$, and go STEP I1.

**STEP I3.**   If $i = (z - 1)$, halt.

**STEP I4.**   If $i = 0$, let $\Xi_z(i) = 0$, $i = 1$, and go to STEP I2.

**STEP I5.**   Let $\Xi_z(i) = \Xi_z(i - 1) + 1$, $i = i + 1$, and go to STEP I2.

The major difference between the proposed implicit BAT and original BAT lies in STEP B2. STEP B2 is changed from "$X(a_i) = 0$" in the BAT to "$\Xi_z(a_i) < \Xi_z(a_{i+1}) - 1$." Therefore, the sequence of $\Xi_z$ is

$$\Xi_z(i) < \Xi_z(j) \text{ if } i < j, \tag{3}$$

$$X_i \ll X_j \text{ if } i < j. \tag{4}$$

Therefore, it is impossible to have a duplicate $X$. For example, two different indicator vectors (0, 1, 1) and (0, 0, 1) indicate the same state vector $X = (1, 1, 0, 0, 0)$ in Fig. 1. Additionally, the BAT is an efficient implicit enumeration, meaning all $\Xi_z$ are found with no duplicates. Therefore, each $X$ with One($X$) = $z$ is found without duplicates because there is a one-to-one relationship between $\Xi_z$ and $X$. Therefore, the algorithm above is correct.

The time complexity of the proposed implicit BAT depends on the number of vectors obtained (i.e., |{$X$ | for all $X$ with One($X$) = $z$}| = {$\Xi_z$ | for all $\Xi_z$}). Because |{$X$ | for all $X$ with One($X$) = $z$}| = $\binom{m}{z}$, the time complexity is $O(\binom{m}{z})$ for the proposed implicit BAT.

**4.3 Example**

Based on the pseudocode for the proposed implicit BAT, we have all the vectors $X$ with One($X$) = $z$ = 3 in Fig. 1 below.

**STEP I0.** Let $i = 0$ and $\Xi_3 = (0, 1, 2)$.

**STEP I1.** Based on $\Xi_3 = (0, 1, 2)$, we have the corresponding $X = (1, 1, 1, 0, 0)$ such that $X(a_j) = 1$ if $\Xi_z(i) = j$.

**STEP I2.** Because $\Xi_z(0) = 0 = \Xi_z(1) - 1$, go to STEP I3.

**STEP I3.** Because $i = 0 < (z - 1) = 2$, go to STEP I4.

**STEP I4.** Because $i = 0$, let $\Xi_z(i) = 0$, $i = 1$, and go to STEP I2.

**STEP I2.** Because $\Xi_z(1) = \Xi_z(2) - 1 = 1$, go to STEP I3.

**STEP I3.** Because $i = 1 < (z - 1) = 2$, go to STEP I4.

**STEP I4.** Because $i > 0$, go to STEP I5.

**STEP I5.** Let $i = i + 1 = 2$ and go to STEP I2.

**STEP I2.** Because $\Xi_z(2) = 2 < \Xi_z(3) - 1 = 4$, let $\Xi_z(2) = \Xi_z(2) + 1 = 3$, $i = 0$, and go STEP I1.

**STEP I1.** Based on $\Xi_3 = (0, 1, 3)$, we have the corresponding $X = (1, 1, 0, 1, 0)$ such that $X(a_j) = 1$ if $\Xi_z(i) = j$.

**STEP I2.** Because $\Xi_z(0) = \Xi_z(1) - 1 = 0$, go to STEP I3.

**STEP I3.** Because $i = 0 < (z - 1) = 2$, go to STEP I4.

**STEP I4.** Because $i = 0$, let $\Xi_z(i) = 0$, $i = 1$, and go to STEP I2.

**STEP I2.** Because $\Xi_z(1) = 1 < \Xi_z(2) - 1 = 2$, let $\Xi_z(2) = \Xi_z(2) + 1 = 3$, $i = 0$, and go STEP I1.

**STEP I1.** Based on $\Xi_3 = (0, 2, 3)$, we have the corresponding $X = (1, 0, 1, 1, 0)$ such that $X(a_j) = 1$ if $\Xi_z(i) = j$.

$$\vdots$$

**STEP I2.** Because $\Xi_z(2) = \Xi_z(3) - 1 = 4$, go to STEP I1.

**STEP I3.** Because $i = (z-1) = 2$, halt.

Based on the procedure above, we can obtain all the vectors $X$ with One($X$) = $z$ = 3. Additionally, in Fig. 1, the procedures for obtaining all $X$ with One($X$) = 1, 2, 3, and 4 are listed in Table 4.

**Table 4.** All $X$ with One($X$) = 1, 2, 3, 4 obtained from the implicit BAT based on Fig. 1.

| iteration | $\Xi_z$ One($X$)=4 | $\Xi_z$ One($X$)=3 | $\Xi_z$ Zero($X$)=2 | $\Xi_z$ One($X$)=1 |
|---|---|---|---|---|
| 1 | (0, 1, 2, 3) (1, 1, 1, 1, 0) | (0, 1, 2) (1, 1, 1, 0, 0) | (0, 1) (1, 1, 0, 0, 0) | (0) (1, 0, 0, 0, 0) |
| 2 | (0, 1, 2, 4) (1, 1, 1, 0, 1) | (0, 1, 3) (1, 1, 0, 1, 0) | (0, 2) (1, 0, 1, 0, 0) | (1) (0, 1, 0, 0, 0) |
| 3 | (0, 1, 3, 4) (1, 1, 0, 1, 1) | (0, 2, 3) (1, 0, 1, 1, 0) | (1, 2) (0, 1, 1, 0, 0) | (2) (0, 0, 1, 0, 0) |
| 4 | (0, 2, 3, 4) (1, 0, 1, 1, 1) | (1, 2, 3) (0, 1, 1, 1, 0) | (0, 3) (1, 0, 0, 1, 0) | (3) (0, 0, 0, 1, 0) |
| 5 | (1, 2, 3, 4) (0, 1, 1, 1, 1) | (0, 1, 4) (1, 1, 0, 0, 1) | (1, 3) (0, 1, 0, 1, 0) | (4) (0, 0, 0, 0, 1) |
| 6 | | (0, 2, 4) (1, 0, 1, 0, 1) | (2, 3) (0, 0, 1, 1, 0) | |
| 7 | | (1, 2, 4) (0, 1, 1, 0, 1) | (0, 4) (1, 0, 0, 0, 1) | |
| 8 | | (0, 3, 4) (1, 0, 0, 1, 1) | (1, 4) (0, 1, 0, 0, 1) | |
| 9 | | (1, 3, 4) (0, 1, 0, 1, 1) | (2, 4) (0, 0, 1, 0, 1) | |
| 10 | | (2, 3, 4) (0, 0, 1, 1, 1) | (3, 4) (0, 0, 0, 1, 1) | |

**4.4 Relationship between Zero($X$) and One($X$)**

Because Zero($X$) + One($X$) = $m$, there are some interesting phenomena between zero ($X$) and one ($X$).

1. $\{X \mid \text{One}(X) = z\} = \{X \mid \text{Zero}(X) = (m - z)\}$

2. The $i$th vector in $\{X \mid \text{One}(X) = z\}$ is equal to the $(|\Omega_z| - i)$th vector in $\Omega_z = \{X \mid \text{Zero}(X) = m-z\}$.

3. $\{\Xi_z^*(i) \mid \text{for all } i\} \cup \{\Xi_z(i) \mid \text{for all } i\} = \{0, 1, \ldots, (m - 1)\}$ and $\{\Xi_z^*(i) \mid \text{for all } i\} \cap \{\Xi_z(i) \mid \text{for all } i\} = \varnothing$.

For Fig. 1, the obtained vectors $X$ correspond to Zero($X$) = 1, 2, 3, 4. These results are presented in Table 5. According to Tables 4 and 5, the three observations above hold true. For example, each state vector $X$ with One($X$) = 3 is also with Zero($X$) = ($m$−3) = 2. For example, there are 10 vectors with One($X$) = 2 and Zero($X$) = 3. The obtained sequence for the vectors with One($X$) = 3 is in the reverse order of that for the vectors with Zero($X$) = 2. For example, $X$ = (0, 0, 1, 1, 1) is the last vector with One($X$) = 3 in Table 4, but is also the first vector with Zero($X$) = 2 in Table 5. Furthermore, the indicator vectors of $X$ = (0, 0, 1, 1, 1) are (2, 3, 4) and (0, 1) in Tables 3 and 4, respectively, and {2, 3, 4} ∪ {0, 1} = {0, 1, …, 4} and {2, 3, 4} ∩ {0, 1} = ∅.

**Table 5.** All $X$ with Zero($X$) = 1, 2, 3, 4 in the proposed implicit BAT based on Fig. 1.

|    | $\Xi_z^*$ | Zero($X$)=1    | $\Xi_z^*$ | Zero($X$)=2     | $\Xi_z^*$  | Zero($X$)=3     | $\Xi_z^*$     | Zero($X$)=4     |
|----|-----------|----------------|-----------|-----------------|------------|-----------------|---------------|-----------------|
| 1  | (0)       | (0, 1, 1, 1, 1)| (0, 1)    | (0, 0, 1, 1, 1) | (0, 1, 2)  | (0, 0, 0, 1, 1) | (0, 1, 2, 3)  | (0, 0, 0, 0, 1) |
| 2  | (1)       | (1, 0, 1, 1, 1)| (0, 2)    | (0, 1, 0, 1, 1) | (0, 1, 3)  | (0, 0, 1, 0, 1) | (0, 1, 2, 4)  | (0, 0, 0, 1, 0) |
| 3  | (2)       | (1, 1, 0, 1, 1)| (1, 2)    | (1, 0, 0, 1, 1) | (0, 2, 3)  | (0, 1, 0, 0, 1) | (0, 1, 3, 4)  | (0, 0, 1, 0, 0) |
| 4  | (3)       | (1, 1, 1, 0, 1)| (0, 3)    | (0, 1, 1, 0, 1) | (1, 2, 3)  | (1, 0, 0, 0, 1) | (0, 2, 3, 4)  | (0, 1, 0, 0, 0) |
| 5  | (4)       | (1, 1, 1, 1, 0)| (1, 3)    | (1, 0, 1, 0, 1) | (0, 1, 4)  | (0, 0, 1, 1, 0) | (1, 2, 3, 4)  | (1, 0, 0, 0, 0) |
| 6  |           |                | (2, 3)    | (1, 1, 0, 0, 1) | (0, 2, 4)  | (0, 1, 0, 1, 0) |               |                 |
| 7  |           |                | (0, 4)    | (0, 1, 1, 1, 0) | (1, 2, 4)  | (1, 0, 0, 1, 0) |               |                 |
| 8  |           |                | (1, 4)    | (1, 0, 1, 1, 0) | (0, 3, 4)  | (0, 1, 1, 0, 0) |               |                 |
| 9  |           |                | (2, 4)    | (1, 1, 0, 1, 0) | (1, 3, 4)  | (1, 0, 1, 0, 0) |               |                 |
| 10 |           |                | (3, 4)    | (1, 1, 1, 0, 0) | (2, 3, 4)  | (1, 1, 0, 0, 0) |               |                 |

## 5. PROPOSED AppBAT

The procedure for the proposed AppBAT for generating a certain number of vectors for approximating binary-state network reliability according to the discussions in Section 4 is presented in Section 5.1. Section 5.2 demonstrates how to apply the proposed AppBAT using an example. Furthermore, the relationships between the approximate reliability $R_z$, arc reliability Pr($a$) = $p$ for all $an \in E$, number of arcs $m$, and z are discussed in Section 5.3.

### 5.1 Pseudocode and Time Complexity of AppBAT

The pseudocode for the proposed AppBAT for calculating the approximate reliability of binary-state networks is presented below.

**Algorithm: AppBAT**

**Input:**  A binary-state network $G(V, E, \mathbf{D})$ and $z$.

**Output:**  The approximate reliability $R_z$.

**STEP 0.**  Let $R_z = 0$ and $\zeta = z$.

**STEP 1.**  Let $i = 0$, $\Xi_\zeta = (0, 1, 2, …, \zeta–1)$, and $\Xi_\zeta(\zeta) = m$.

**STEP 2.** Use $\Xi_\zeta$, to find the corresponding $X$ such that $X(a_j) = 1$ if $\Xi_\zeta(i) = j$ for one $i$ and $X(a_j) = 0$ for all $i$.

**STEP 3.** Let $R_z = R_z + \Pr(X)$ if $X$ is connected after verifying using the PLSA.

**STEP 4.** If $\Xi_\zeta(i) + 1 < \Xi_\zeta(i + 1)$, let $\Xi_\zeta(i) = \Xi_\zeta(i) + 1$, $i = 0$, and go to STEP 2.

**STEP 5.** If $i = (\zeta - 1)$, go to STEP 8.

**STEP 6.** If $i = 0$, let $\Xi_\zeta(0) = 0$, $i = 1$, and go to STEP 4.

**STEP 7.** Let $\Xi_\zeta(i) = \Xi_\zeta(i - 1) + 1$, $i = i + 1$, and go to STEP 4.

**STEP 8.** If $\zeta < m$, let $\zeta = \zeta + 1$ and go to STEP 1. Otherwise, halt.

Loop STEPs 1 to 7 are based on the proposed implicit BAT with a time complexity of $O(\binom{m}{\zeta})$. In the loop from STEPs 1 to 7, STEP 3 verifies the connectivity of $X$ using the PLSA with a time complexity of $O(n)$. There are at least $(m - z + 1)$ loops. Therefore, the total time complexity of the proposed AppBAT is $O(n(m - z + 1)\binom{m}{\zeta})$, which is less than that of the BAT, which is $O(2^{m+1})$. Therefore, the proposed AppBAT is more efficient than the BAT. Time complexity is always effective for comparing the performance of related network reliability algorithms.

**5.2 Example**

To demonstrate the step-by-step procedure of the proposed AppBAT for finding each connected vector in $\Omega_3 = \{X \mid X \text{ is connected and } 3 \leq \text{One}(X)\}$ and calculate the approximated reliability $R_3 = \Pr(\Omega_3)$ for all arcs $a$ with $\Pr(a) = 0.9$, we implemented the binary-state network shown in Fig. 1.

**Solution:**

**STEP 0.** Let $R_z = 0$ and $\zeta = 3$.

**STEP 1.** Let $i = 0$, $\Xi_\zeta = (0, 1, 2)$, and $\Xi_\zeta(3) = 5$.

**STEP 2.** The corresponding state vector $X$ for $\Xi_\zeta = (0, 1, 2)$ is $(1, 1, 1, 0, 0)$, $X_\zeta(a_0) = 0 < X_\zeta(a_1) - 1 = 2$, $X_\zeta(a_0) = X_\zeta(a_0) + 1 = 1$, $i = 0$, and go to STEP 3.

**STEP 3.** Because $X$ is disconnected, go to STEP 4.

**STEP 4.** Because $\Xi_\zeta(0) + 1 = \Xi_\zeta(1) = 1$, go to STEP 5.

**STEP 5.** If $i = 0 < (\zeta - 1) = 2$, go to STEP 6.

**STEP 6.** Because $i = 0$, let $\Xi_\zeta(0) = 0$, $i = 1$, and go to STEP 4.

**STEP 4.** Because $\Xi_\zeta(1) + 1 = \Xi_\zeta(2) = 2$, go to STEP 5.

**STEP 5.** If $i = 1 < (\zeta - 1) = 2$, go to STEP 6.

**STEP 6.** Because $i = 1 > 0$, go to STEP 7.

**STEP 7.** Let $\Xi_\zeta(1) = \Xi_\zeta(0) + 1 = 1$, $i = i + 1 = 2$, and go to STEP 4.

**STEP 4.** Because $\Xi_\zeta(2) + 1 = 3 < \Xi_\zeta(3) = 5$, let $\Xi_\zeta(2) = \Xi_\zeta(2) + 1 = 3$, $i = 0$, and go to STEP 2.

**STEP 2.** The corresponding state vector $X$ for $\Xi_\zeta = (0, 1, 3)$ is $(1, 1, 0, 1, 0)$ and go to STEP 3.

**STEP 3.** Because $X$ is connected, let $R_z = R_z + \Pr(X) = 0.007290$ and go to STEP 4.

$$\vdots$$

**STEP 4.** Because $\Xi_\zeta(2) + 1 = \Xi_\zeta(3) = 5$, go to STEP 5. Note that $\Xi_\zeta = (2, 3, 4)$.

**STEP 5.** Because $i = (\zeta - 1) = 2$, go to STEP 8.

**STEP 8.** Because $\zeta = 3 < m = 5$, let $\zeta = \zeta + 1 = 4$ and go to STEP 1.

**STEP 1.** Let $i = 0$, $\Xi_\zeta = (0, 1, 2, 3)$, and $\Xi_\zeta(4) = 5$.

**STEP 2.** The corresponding state vector for $\Xi_\zeta = (0, 1, 2, 3)$ is $X = (1, 1, 1, 1, 0)$ and go to STEP 3.

**STEP 3.** Because $X$ is connected, let $R_z = R_z + \Pr(X) = 0.05103 + 0.065610 = 0.11664$ and go to STEP 4.

$$\vdots$$

**STEP 4.** Because $\Xi_\zeta(3) + 1 = \Xi_\zeta(4) = 5$, go to STEP 5. Note that $\Xi_\zeta = (1, 2, 3, 4)$.

**STEP 5.** Because $i = (\zeta - 1) = 3$, go to STEP 8.

**STEP 8.** Because $\zeta = 4 < m = 5$, let $\zeta = \zeta + 1 = 5$ and go to STEP 1.

**STEP 1.** Let $i = 0$, $\Xi_\zeta = (0, 1, 2, 3, 4)$, and $\Xi_\zeta(5) = 5$.

**STEP 2.** The corresponding state vector for $\Xi_\zeta = (0, 1, 2, 3, 4)$ is $X = (1, 1, 1, 1, 1)$ and go to STEP 3.

**STEP 3.** Because $X$ is connected, let $R_z = R_z + \Pr(X) = 0.24786 + 0.590490 = 0.96957$ and go to STEP 4.

$$\vdots$$

**STEP 4.** Because $\Xi_\zeta(4) + 1 = \Xi_\zeta(5) = 5$, go to STEP 5. Note that $\Xi_\zeta = (0, 1, 2, 3, 4)$.

**STEP 5.** Because $i = (\zeta - 1) = 4$, go to STEP 8.

**STEP 8.** Because $\zeta = m = 5$, halt.

The final results of the proposed AppBAT for calculating the approximated reliability $R_z = \Pr(\Omega_z)$, where $\Omega_z = \{X \mid X \text{ is connected and } \text{Zero}(X) \leq z\}$, are listed in Table 6.

Table 6. Final results of the proposed AppBAT for $z = 3$ in Fig. 1.

| One($X_i$) | iteration | $X$ | Connected? | $\Pr_{0.99}(X)$ | $\Pr_{0.9}(X)$ | $\Pr_{0.8}(X)$ | $\Pr_{0.5}(X)$ |
|---|---|---|---|---|---|---|---|
| 3 | 1 | (1, 1, 1, 0, 0) | N | | | | |
| | 2 | (1, 1, 0, 1, 0) | Y | 0.000097 | 0.007290 | 0.020480 | 0.031250 |
| | 3 | (1, 0, 1, 1, 0) | Y | 0.000097 | 0.007290 | 0.020480 | 0.031250 |
| | 4 | (0, 1, 1, 1, 0) | N | | | | |
| | 5 | (1, 1, 0, 0, 1) | Y | 0.000097 | 0.007290 | 0.020480 | 0.031250 |
| | 6 | (1, 0, 1, 0, 1) | Y | 0.000097 | 0.007290 | 0.020480 | 0.031250 |
| | 7 | (0, 1, 1, 0, 1) | Y | 0.000097 | 0.007290 | 0.020480 | 0.031250 |
| | 8 | (1, 0, 0, 1, 1) | Y | 0.000097 | 0.007290 | 0.020480 | 0.031250 |
| | 9 | (0, 1, 0, 1, 1) | Y | 0.000097 | 0.007290 | 0.020480 | 0.031250 |
| | 10 | (0, 0, 1, 1, 1) | | | | | |
| 4 | 11 | (1, 1, 1, 1, 0) | Y | 0.009606 | 0.065610 | 0.081920 | 0.031250 |
| | 12 | (1, 1, 1, 0, 1) | Y | 0.009606 | 0.065610 | 0.081920 | 0.031250 |
| | 13 | (1, 1, 0, 1, 1) | Y | 0.009606 | 0.065610 | 0.081920 | 0.031250 |
| | 14 | (1, 0, 1, 1, 1) | Y | 0.009606 | 0.065610 | 0.081920 | 0.031250 |
| | 15 | (0, 1, 1, 1, 1) | Y | 0.009606 | 0.065610 | 0.081920 | 0.031250 |
| 5 | 16 | (1, 1, 1, 1, 1) | Y | 0.950990 | 0.590490 | 0.327680 | 0.031250 |
| | SUM | | | 0.999699 | 0.969570 | 0.880640 | 0.406250 |
| | Exact | | | 0.999701 | 0.971190 | 0.890880 | 0.468750 |

In Table 6, one can see that the larger the component reliability $\Pr(a)$, the better the approximate reliability $R_z$. The results obtained after fixing the value of component reliability are presented in Table 7, where $X_i$ is the $i$th state vector listed in Table 2. The vectors with strikethrough are disconnected (e.g., $X_{15}$). In Table 7, it is clear that $R_i < R_j$ if $i < j$.

Table 7. Final results of the proposed AppBAT for Fig. 1 at $p = 0.9$.

| $z$ | $\Omega_z$ | | | | | $|\Omega_z|$ | $|\Omega_z|/32$ | $\Pr(\Omega_{z,c})$ | $R-\Pr(\Omega_{z,c})/R$ |
|---|---|---|---|---|---|---|---|---|---|

| 0 | {$X_{32}$} | 1 | 96.88% | 0.59049 | 39.20% |
|---|---|---|---|---|---|
| 1 | {$X_{16}$, $X_{24}$, $X_{28}$, $X_{30}$, $X_{31}$, $X_{32}$} | 6 | 81.25% | 0.91854 | 5.42% |
| 2 | {$X_{12}$, $X_{14}$, $\cancel{X}_{15}$, $X_{16}$, $X_{20}$, $X_{22}$, $X_{23}$, $X_{24}$, $X_{26}$, $X_{27}$, $X_{28}$, $\cancel{X}_{29}$, **$X_{30}$**, **$X_{31}$**, **$X_{32}$**} | 15 | 53.13% | 0.96957 | 0.17% |
| 3 | {$X_{10}$, $\cancel{X}_{11}$, $X_{12}$, $\cancel{X}_{13}$, $X_{14}$, $\cancel{X}_{15}$, $X_{16}$, $\cancel{X}_{18}$, $X_{19}$, $X_{20}$, $\cancel{X}_{21}$, $X_{22}$, $X_{23}$, $X_{24}$, **$X_{25}$**, $X_{26}$, $X_{27}$, $X_{28}$, $\cancel{X}_{29}$, **$X_{30}$**, **$X_{31}$**, **$X_{32}$**} | 22 | 31.25% | 0.97119 | 0.00% |
| 4 | {$X_{10}$, $\cancel{X}_{11}$, $X_{12}$, $\cancel{X}_{13}$, $X_{14}$, $\cancel{X}_{15}$, $X_{16}$, $\cancel{X}_{17}$, $\cancel{X}_{18}$, $X_{19}$, $X_{20}$, $\cancel{X}_{21}$, $X_{22}$, $X_{23}$, $X_{24}$, **$X_{25}$**, $X_{26}$, $X_{27}$, $X_{28}$, $\cancel{X}_{29}$, **$X_{30}$**, **$X_{31}$**, **$X_{32}$**} | 23 | 28.13% | 0.97119 | 0.00% |
| 5 | {$X_{10}$, **$X_{11}$**, $X_{12}$, **$X_{13}$**, $X_{14}$, **$X_{15}$**, $X_{16}$, **$X_{17}$**, **$X_{18}$**, $X_{19}$, $X_{20}$, **$X_{21}$**, $X_{22}$, $X_{23}$, $X_{24}$, **$X_{25}$**, $X_{26}$, $X_{27}$, $X_{28}$, **$X_{29}$**, $\cancel{X}_{30}$, $\cancel{X}_{31}$, $\cancel{X}_{32}$} | 23 | 28.13% | 0.97119 | 0.00% |

### 5.3 Computation Experiments

The practical performance of the proposed AppBAT was further tested on two large-scale benchmark networks, as shown in Fig. 3.

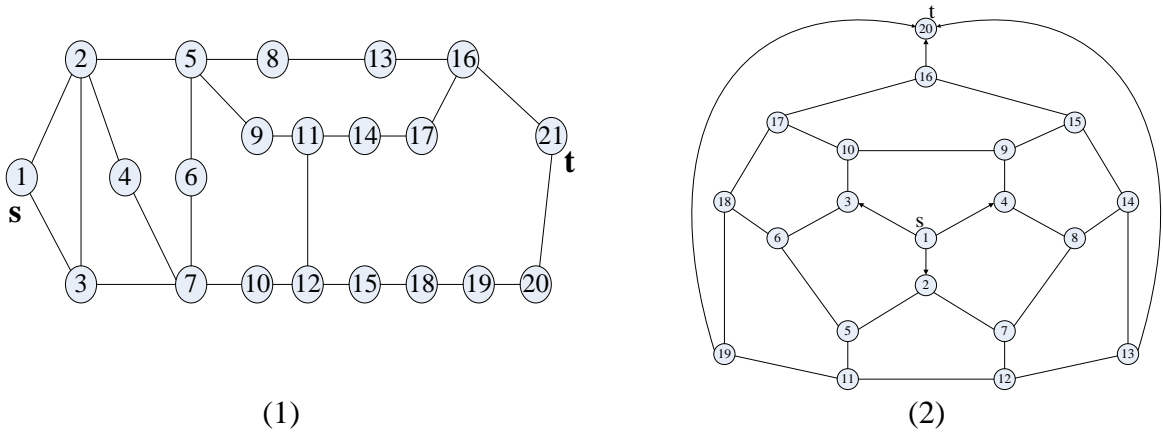

(1)             (2)
**Figure 3.** Two large-scale benchmark binary-state networks.

These two networks have been extensively implemented in many studies to verify the effectiveness and efficacy of new algorithms. In Fig. 3(1), we have $n = 21$, $m = 48$, $n_p = 6$, $n_c = 2$, $|P| = 44$, and the maximum MP-based combinations on $2^{|P|} = 1.75922\text{E}+13$ [52]. in Fig. 3(2), we have $n = 20$, $m = 54$, $n_p = 5$, $n_c = 3$, and $|P|$ is unknown.

The proposed AppBAT was coded in DEV C++ and executed on a notebook with an Intel Core i7-10750H CPU @2.60 GHz with 64 GB of memory (64 bit Windows 10). Three cases were considered based on the value of highly reliable components, namely $\Pr(a) = 0.99$, 0.95, and 0.9, for all arcs $a$ for each test benchmark network. AppBAT is forced to terminate if its runtime exceeds one hour when calculating $R_z$ for $z = m, m - 1, \ldots$. The results include $|\overline{\Omega}_z|$, $|\Omega_z|$, $\Pr_p(\Omega_z)$, and $T_p(\Omega_z)$ for $p = 0.99$, 0.95, and 0.9 with $z = 48–37$ for Fig. 3(1) and $z = 54–44$ for Fig. 3(2).

**Table 8.** Results for Fig. 3(1).

| z | $|\bar{\Omega}_z|$ | $|\Omega_z|$ | $\Pr(\Omega_{z,0.99})$ | $T(\Omega_{z,0.99})$ | $\Pr(\Omega_{z,0.95})$ | $T(\Omega_{z,0.95})$ | $\Pr(\Omega_{z,0.90})$ | $T(\Omega_{z,0.90})$ |
|---|---|---|---|---|---|---|---|---|
| 48 | 1 | 1 | 0.617290440 | 0.000 | 0.08525754 | 0.000 | 0.006362680 | 0.000 |
| 47 | 48 | 48 | 0.299292030 | 0.000 | 0.21538752 | 0.000 | 0.033934290 | 0.000 |
| 46 | 1128 | 1122 | 0.070666100 | 0.000 | 0.26498339 | 0.000 | 0.088134910 | 0.000 |
| 45 | 17296 | 16964 | 0.010792220 | 0.007 | 0.21086310 | 0.006 | 0.148061100 | 0.006 |
| 44 | 194580 | 186039 | 0.001195500 | 0.071 | 0.12170903 | 0.056 | 0.180415690 | 0.059 |
| 43 | 1712304 | 1574485 | 0.000102200 | 0.508 | 0.05421305 | 0.480 | 0.169654920 | 0.471 |
| 42 | 12271512 | 10690501 | 0.000007010 | 3.384 | 0.01937358 | 3.410 | 0.127992230 | 3.234 |
| 41 | 73629072 | 59806870 | 0.000000400 | 19.334 | 0.00570439 | 19.594 | 0.079559900 | 18.725 |
| 40 | 377348994 | 281085174 | 0.000000020 | 98.307 | 0.00141105 | 102.804 | 0.041546910 | 99.265 |
| 39 | 1677106640 | 1126417710 | 0.000000000 | 433.948 | 0.00029761 | 449.710 | 0.018499410 | 433.859 |
| 38 | 6540715896 | 3894214144 | 0.000000000 | 1633.911 | 0.00005415 | 1705.260 | 0.007106170 | 1674.573 |
| 37 | 22595200368 | 3135194112 | 0.000000000 | 5473.193 | 0.00000229 | 5423.388 | 0.000635680 | 5539.216 |
| $R_{40}$ | | | 0.999345920 | 122.611 | 0.978902650 | 126.350 | 0.875662630 | 121.760 |
| $R_{39}$ | | | 0.999345920 | 122.611 | 0.979200260 | 576.060 | 0.894162040 | 555.619 |
| $R_{38}$ | | | 0.999345920 | 122.611 | 0.979254410 | 2281.320 | 0.901268210 | 2230.192 |
| $R_{37}$ | | | 0.999345920 | 122.611 | 0.979256700 | 8004.708 | 0.901903890 | 7769.408 |

**Table 9.** Results for Fig. 3(2).

| z | $|\bar{\Omega}_z|$ | $|\Omega_z|$ | $\Pr(\Omega_{z,0.99})$ | $T(\Omega_{z,0.99})$ | $\Pr(\Omega_{z,0.95})$ | $T(\Omega_{z,0.95})$ | $\Pr(\Omega_{z,0.90})$ | $T(\Omega_{z,0.90})$ |
|---|---|---|---|---|---|---|---|---|
| 54 | 1 | 1 | 0.581166740 | 0.000 | 0.062672120 | 0.000 | 0.003381390 | 0.000 |
| 53 | 54 | 54 | 0.316999730 | 0.000 | 0.178120810 | 0.000 | 0.020288330 | 0.000 |
| 52 | 1431 | 1431 | 0.084853380 | 0.001 | 0.248431710 | 0.000 | 0.059737870 | 0.000 |
| 51 | 24804 | 24802 | 0.014855270 | 0.009 | 0.226621240 | 0.009 | 0.115041460 | 0.009 |
| 50 | 316251 | 316143 | 0.001912680 | 0.105 | 0.152035150 | 0.108 | 0.162932930 | 0.106 |
| 49 | 3162510 | 3159642 | 0.000193090 | 1.006 | 0.079973270 | 1.102 | 0.180934250 | 1.032 |
| 48 | 25827165 | 25777194 | 0.000015910 | 7.887 | 0.034339120 | 8.149 | 0.164012210 | 8.204 |
| 47 | 177100560 | 176457504 | 0.000001100 | 53.709 | 0.012372000 | 54.750 | 0.124749340 | 55.072 |
| 46 | 1040465790 | 1033944255 | 0.000000070 | 315.765 | 0.003815430 | 316.301 | 0.081218110 | 316.326 |
| 45 | 5317936260 | 968682204 | 0.000000000 | 1579.972 | 0.000188140 | 1685.585 | 0.008454630 | 1580.875 |
| 44 | 23930713170 | 2074672378 | 0.000000000 | 6945.425 | 0.000021210 | 7369.413 | 0.002011970 | 6946.389 |
| $R_{47}$ | | | 0.999997900 | 62.717 | 0.994565420 | 64.118 | 0.831077780 | 64.423 |
| $R_{46}$ | | | 0.999997970 | 378.482 | 0.998380850 | 380.419 | 0.912295890 | 380.749 |
| $R_{45}$ | | | 0.999997970 | 1958.454 | 0.998568990 | 2066.004 | 0.920750520 | 1961.624 |
| $R_{44}$ | | | 0.999997970 | 8903.879 | 0.998590200 | 9435.417 | 0.922762490 | 8908.013 |

Based on the experimental results for Fig. 3 in Tables 8 and 9, we can make the following observations.

1. The smaller the value of $z$, the greater the value of $\sum_{i=z}^{m}|\bar{\Omega}_i|$, the greater the value of $\sum_{i=z}^{m}|\Omega_i|$, and the better the value of $R_z = \sum_{i=z}^{m} \Pr(\Omega_i)$. This observation can be used to define a stopping criterion to determine the value of $z$ when finding $R_z$. For example, one could set the value of $z$ such that $|R_z| - |R_{z+1}| = \Pr_p(\Omega_z) < 10^{-5}$.

2. $R_{z,p} > R_{z,p^*}$ if $p > p^*$, meaning the proposed AppBAT is more accurate for networks with a larger component reliability.

3. $|\bar{\Omega}_{z,p}| = |\bar{\Omega}_{z,p^*}|$, $|\Omega_{z,p}| = |\Omega_{z,p^*}|$, and $T(\Omega_{z,p}) \approx T(\Omega_{z,p^*})$, regardless of the values of $p$ and $p^*$.

4. The values of $\Pr(\Omega_z)$ increase with a decrease in $z$ until reaching the maximum, and then start to decrease with a decrease in $z$. For example, for $p = 0.90$, the value of $\Pr(\Omega_z)$ increases from $z = m = 54$ to 50, and then decreases from $z = 48$ to 44 after reaching a maximum at $z = 49$, as shown in Table 9.

5. Let $z_p$ be a value such that $\Pr\left(\Omega_{z_p}\right) \geq \Pr(\Omega_z)$ for all $z$. We have $\Pr\left(\Omega_{z_{p*}}\right) < \Pr\left(\Omega_{z_p}\right)$ and $z_{p*} < z_p$ if $p^* < p$. For example, $z_p = 54$ and 49 for $p = 0.99$ and 0.9, respectively. We have $\Pr\left(\Omega_{z_{p,0.99}}\right) = 0.581166740 > \Pr\left(\Omega_{z_{p,0.90}}\right) = 0.18093425$ and $54 > 49$. Therefore, we can improve the accuracy of the approximate reliability if we can find $\Omega_z$ such that $z$ is equal to or close to $z_p$. Additionally, this observation explains Observation 4 and confirms that the proposed AppBAT is able to provide good approximate reliability for a highly reliable network.

6. Because $|\overline{\Omega}_z| = \binom{m}{z}$, $|\overline{\Omega}_z|$ increases with increasing values of One($X$) for One($X$) < $m$ / 2, reaches a maximum for One($X$) = $m$ / 2, and then decreases with an increase in One($X$) for $m$ / 2 < One($X$). This observation can help us determine the correct value of $z$ for deriving a better $R_z$.

7. The trend in the number of connected state vectors mimics the trend in the total number of state vectors, meaning the more state vectors, the greater the number of connected state vectors in general. However, there are some special scenarios that do not follow this general observation. For example, the numbers of connected vectors are 1033944255, 968682204, and 2074672378 for One($X$) = 46, 45, and 44, respectively, but the number of state vectors increases from One($X$) = 46 to One($X$) = 45 and to One($X$) = 44.

8. The greater the Pr($a$) for all $an \in E$, the greater the number of $z$ and the shorter the runtime required to obtain a better $R_z$. This observation can also be applied to any large-scale network to derive a better approximate reliability in a short time.

9. The runtime required to derive all connected vectors $X$ is exponential relative to the value of One($X$) based on the characteristic of NP-hardness. More importantly, the runtime is fixed with the number of state vectors and not the number of connected state vectors. Therefore, the PLSA is very efficient at verifying connectivity and its runtime can be ignored. This observation can help to set a time limit for calculating approximate reliability (e.g., 3600 s in our experiments). This further explains the efficiency issue in the third observation.

Based on experimental results presented above, the proposed AppBAT based on the implicit BAT can calculate the approximate reliability $R_z$ for large-scale, highly reliable binary-state networks according to the required runtime and the difference between $R_z$ and $R_{z+1}$. These results reinforce the inferences based on time complexity presented in Section 5.1.

## 6. CONCLUSIONS

This paper presented a novel and simple direct approximate reliability method called AppBAT without using either the MCS or MPs/MCs to calculate the approximate reliability of large-scale, highly reliable binary-state networks.

The basic concept of the proposed AppBAT is to count only connected vectors with $z \leq One(X)$ to obtain an approximate $R_z$, rather than counting all connected vectors like the traditional BAT. The proposed AppBAT uses the PLSA to verify the connectivity of the identified vectors and the proposed novel implicit BAT to find only vectors $X$ with $One(X) = z$ without needing to have all state vectors.

Computational experiments on two large-scale benchmark problems revealed that the proposed AppBAT is able to derive good approximate reliability values within 10 min, particularly for highly reliable networks with $Pr(a) = 0.99$. There has been no report of calculating the exact reliability of the two benchmarks within 10 min using a typical notebook computer. Therefore, from a computational, pragmatic, and theoretical perspective, the proposed AppBAT is more attractive than existing exact reliability algorithms, MCS-based algorithms, and reliability bound algorithms for calculating the reliability of highly reliable binary-state networks.

**ACKNOWLEDGMENT**

This research was supported in part by the Ministry of Science and Technology, R.O.C. under grant MOST 107-2221-E-007-072-MY3 and MOST 110-2221-E-007-107-MY3. This article was once submitted to arXiv as a temporary submission that was just for reference and did not provide the copyright.